# ReuNify: A Step Towards Whole Program Analysis for React Native Android Apps

Yonghui Liu[†], Xiao Chen[†*], Pei Liu[‡], John Grundy[†], Chunyang Chen[†], Li Li[§*]
[†] *Monash University*, Melbourne, Australia
[‡] *CSIRO's Data61*, Melbourne, Australia
[§] *Beihang University*, Beijing, China
{Yonghui.Liu, Xiao.Chen, John.Grundy, Chunyang.Chen}@monash.edu,
{Pei.Liu}@data61.csiro.au, {lilicoding}@ieee.org

*Abstract*—React Native is a widely-used open-source framework that facilitates the development of cross-platform mobile apps. The framework enables JavaScript code to interact with native-side code, such as Objective-C/Swift for iOS and Java/Kotlin for Android, via a communication mechanism provided by React Native. However, previous research and tools have overlooked this mechanism, resulting in incomplete analysis of React Native app code. To address this limitation, we have developed REUNIFY, a prototype tool that integrates the JavaScript and native-side code of React Native apps into an intermediate language that can be processed by the Soot static analysis framework. By doing so, REUNIFY enables the generation of a comprehensive model of the app's behavior. Our evaluation indicates that, by leveraging REUNIFY, the Soot-based framework can improve its coverage of static analysis for the 1,007 most popular React Native Android apps, augmenting the number of lines of Jimple code by 70%. Additionally, we observed an average increase of 84% in new nodes reached in the callgraph for these apps, after integrating REUNIFY. When REUNIFY is used for taint flow analysis, an average of two additional privacy leaks were identified. Overall, our results demonstrate that REUNIFY significantly enhances the Soot-based framework's capability to analyze React Native Android apps.

## I. Introduction

Mobile apps have become the primary source of digital consumption, with a growing number of users relying on apps for various purposes such as shopping, entertainment, and communication. As a result, businesses are investing heavily in mobile app development to reach their target audience and remain competitive in the market. Many companies are facing the challenge of needing to build mobile apps for multiple platforms, specifically for both Android and iOS. This *cross-platform mobile app development* has gained popularity due to its consistency across platforms, cost-effectiveness, time efficiency, wide audience reach, and easier maintenance [1].

Nowadays, React Native (used in Facebook, Shopify, Skype, etc.) and Flutter (used in Google Ads, Reflectly, Alibaba, etc.) have become the two most popular frameworks for cross-platform mobile app development [2]. Each of these cross-platform solutions has its own capabilities and strengths.

[*]Xiao Chen and Li Li are the corresponding authors.

React Native, an open-source framework, gained popularity since its 2015 launch by combining traditional mobile development with Node.js-based flexibility. The core idea of React Native is to empower cross-platform JavaScript APIs to invoke platform-specific functions involving invoking Objective-C/Swift or Java/Kotlin functions to utilize iOS and Android components. This feature sets it apart from other cross-platform mobile application development technologies which often end up rendering web-based views. With React Native, developers can create a shared codebase in JavaScript that works on both Android and iOS. This is achieved by providing a set of cross-platform APIs and Components that conceal platform-specific native code and abstract the differences between platforms. React Native is flexible and can be used in existing Android and iOS projects or to create a new app from scratch [3].

The stats from AppBrain [4] report that among the top 500 Android apps in the US, 14.85% of installed apps are built with React Native. In fact, in the category of top 500 US Android apps, React Native is the third most popular framework, right after Kotlin and Android Architecture Components. While the use of the React Native framework can streamline the app development process, it also introduces new challenges for app analysis, particularly in terms of static analysis. The main difficulty with static analysis on React Native apps is their use of multiple programming languages with varying semantics, along with the complex mechanisms inherent in the React Native framework. These factors can make it very challenging to thoroughly analyze and fully comprehend the app's codebase.

In the last decade, Android app analysis has been a prominent research theme in software engineering. Static analysis techniques have been implemented by many approaches and tools for bug detection, security property checking, malware detection, and empirical studies. Unfortunately, as far as we know, there is no existing techniques or tools for analyzing apps developed with React Native. The approaches of the current state-of-the-art tools, which were intended for traditional Android apps, are not sufficient for efficiently covering the executable code in React Native apps due to the complexity of the underlying mechanism of the React Native framework. In light of this challenge, we suggest a new research direction to enable static analysis of the whole program of React Native Android app.

We propose REUNIFY, aiming to fill the gap in the whole-app analysis, by extracting and unifying artefacts from both the Java and JavaScript sides of React Native Android Apps into Jimple [5], the intermediate representation in Soot. To the best of our knowledge, REUNIFY is the first static analyser for React Native Android apps[6]. By transforming JavaScript-side code into Jimple, REUNIFY provides the opportunity for several analyses (e.g., call graph analysis, control flow graph and taint flow analysis) in the literature to readily account for JavaScript code. By modelling React Native mechanism, REUNIFY increases the coverage of Java-side code analysis. REUNIFY is thus a multi-step static analysis approach that we implement as a framework to enable the whole-programme analysis for React Native Android Apps. This research makes the following key contributions:

- We propose REUNIFY, a novel approach to build a unified model of React Native Android app code for facilitating complete static analyses. We have implemented REUNIFY to produce the Jimple code, which facilitates the integration of JavaScript-side code and Dalvik bytecode within a React Native Android app package.
- We show that REUNIFY can significantly enhance React Native Android Apps' call-graphs, revealing previously unreachable methods in React Native Android Apps.
- We evaluate REUNIFY on a set of real-world React Native Android Apps, showing that it enables existing analysers to reveal sensitive data leaks.
- We release our open-source prototype REUNIFY and all artifacts used in our study at:

https://github.com/DannyGooo/ReuNify

The remainder of this paper is organized as follows. We outline the key motivation for this work in Section II, and Section III presents key aspects of our approach. Section IV presents our studied datasets, our experimental setup, and our experimental results. Section V discusses the threats to the validity of our research. Section VI discusses key related work, and Section VII summarises this paper.

## II. BACKGROUND

### A. React Native

React Native is a widely used JavaScript framework that enables developers to build mobile applications for both iOS and Android platforms. It is based on the popular *React* framework [7], which is a Node.js-based JavaScript library used for creating user interfaces. However, JavaScript and React are not natively capable of accessing platform-specific features e.g. Android-specific or iOS-specific features.

As shown in Figure 1, JavaScript can exchange information with the platform side through the underlying mechanism of React Native. As demonstrated in the Figure 1(a), *bridge* [8] was used to facilitate the exchange of information between JavaScript and platform-side code in the old architecture of React Native. *Bridge* allowed JavaScript to interact with the platform-specific features (e.g., Native Components and Native Modules) for building mobile apps. However, this architecture suffered from issues such as asynchronous behavior, single-threading, and extra overheads (JSON format) that impacted performance and flexibility. To address these issues, the new architecture of React Native adopts the *JavaScript Interface* (*JSI*), as shown in Figure 1(b). The *JSI* allows a JavaScript object to hold a reference to a C++ object and vice versa, enabling synchronous execution, concurrency, lower overhead, code sharing, and type safety [8]. This approach provides several advantages over the old architecture and serves as the foundation of the new Native Module System. Using *Turbo Native Modules* and *Fabric Native Components* developers can create high-performance and flexible mobile applications for both iOS and Android platforms.

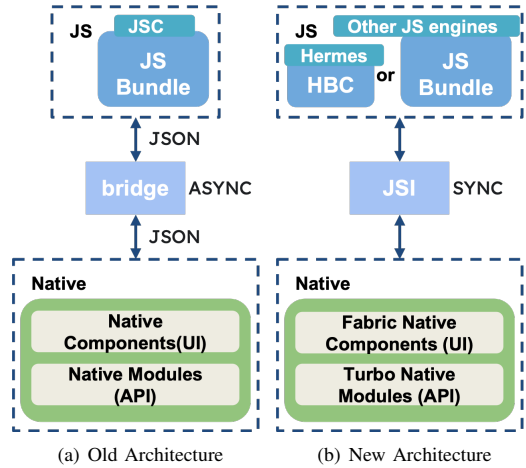

(a) Old Architecture  (b) New Architecture

Fig. 1. Cross-Language Communication Mechanism in React Native

When developing a React Native application, JavaScript is used to organize reusable and nestable *React Components* to implement the user interface. These components can be enhanced with various module APIs to achieve desired features and functionalities. Figure 2 categorizes *React Components* and *Module APIs* based on the entity responsible for maintaining them. React Native comes with built-in core components and APIs that are ready for use [3]. However, developers are not limited to these built-in components and APIs. There are various rich third-party libraries maintained by the community[9]. Apart from the cross-language strategy provided by third-party libraries or React Native's core, developers can wrap their own native-side code to be invoked from the JavaScript side through React Native's underlying cross-language mechanism [10, 11].

### B. JavaScript code in React Native Android APK

When releasing a React Native project as an Android APK, the JavaScript code from React Native is bundled by the *Metro*[12], a JavaScript bundler, which takes in options, an entry file, and gives you a JavaScript file including all JavaScript files back. The JavaScript code inside bundle file is then further compiled into bytecode with Hermes selected as JavaScript Engine. Once the app launches, the code is loaded from the bundled file and executed by the JavaScript engine. The engine runs the code and communicates with the native side through either *bridge* in the old architecture, or *JSI* in the new architecture.

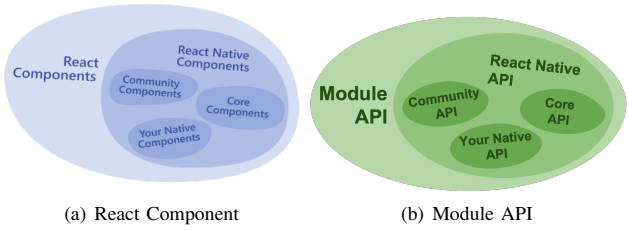

(a) React Component  (b) Module API

Fig. 2. The structure of React Component and Module API in the developer's JavaScript side

Since version 0.70.0 (September 2022), the default JavaScript engine in React Native has been changed from JavaScriptCore (JSC) [13] to Hermes Engine [14]. Before that, Hermes Engine has been introduced to React Native Android and React Native iOS since version 0.60.4 and version 0.64.0 as an optional engine, respectively. The legacy JavaScript Engine parse all JavaScript Codes using just-in-time (JIT) compilation. With the inclusion of Hermes engine, JavaScript source code would be compiled to bytecode ahead of time (AOT), which saves the interpreter from having to perform this expensive step during app startup, and also contributes to a smaller app bundle size. However, the use of the Hermes engine in React Native can make static analysis much more challenging. The generated Hermes bytecode is not as easily readable or accessible as the JavaScript code, which makes the current state-of-the-art tools designed for JavaScript [15, 16, 17] useless in front of Hermes bytecode. Additionally, current state-of-the-art Android static code analysis approaches [18, 19, 6, 20] overlook the apps developed with React Native.

### C. Motivating Example

The React Native framework's complex mechanism conceals a significant portion of the executable code of Android apps built with it from state-of-the-art static analysis tools [19, 21, 22]. With one analysis for the React Native Android Apps, *Skype*, *com.skype.raider* [23], we make the case that React Native mechanism should be considered in static analysis approaches.

*Skype* is a popular app for real-time video calls, with more than one billion installations. This app is developed with React Native framework. Considering the cross-language communication mechanism in React Native, we discuss both the JavaScript side code and Java side code. In the example, we've sourced version 8.83.0.411 of the Skype app from APKMirror[1].

[1] https://www.apkmirror.com/apk/skype/skype-skype/skype-skype-8-83-0-411-release/

**JavaScript Side:** The app, *Skype*, incorporate version 89 of the Hermes engine, and stores the JavaScript-side code as Hermes bytecode. This bytecode can be decompiled into a textual disassembly file containing 3,589,897 lines of text and a file size of 119 megabytes. The file includes 87,400 methods that were not considered in the current research and tools.

**Java Side:** We generated a callgraph of the app, *Skype*, using FlowDroid for taint flow analysis. The callgraph consisted of 5,169 nodes and 18,282 edges, and no privacy leaks were detected. After examining the call graph, it was found that the Java methods exposed through the Native Module API (135 Modules, 724 methods) and React Native Components (106 Components, 813 methods) were not captured in the callgraph. Upon these methods, the call graph expanded considerably to include 13,629 nodes and 51,395 edges, and three privacy leaks were identified.

This paper presents a novel strategy to address the challenge of the hidden executable code in React Native Android apps, which has been a gap in the current research. The aim of this paper is to enable whole program analysis for React Native Android app.

### III. CHALLENGES AND APPROACH

In order to reveal concealed executable code within React Native, we have implemented a prototype, REUNIFY, to enable a more complete static analysis on React Native Android Apps. As depicted in Figure 3, REUNIFY contains two key modules. The first module, *Jimple Code Generation* (1), instantiates the Java-side and JavaScript-side code inside the React Native Android Apps in the Jimple representation (i.e., the intermediate representation in Soot which is the most widely adopted framework for static analysis of Android apps). Then, *Cross-Language Discloser* (2) is implemented based on the pointer analysis on the Jimple Code generated from the first module to detect the statements that are responsible for cross-language communication. We detail the design and implementation of each module in the following subsections. However, due to space constraints, we will not present all technical details related to Jimple. We invite interested readers to consider our open-source project. REUNIFY is fully open-sourced.

### A. Jimple Code Generation

REUNIFY leverages a *divide-and-conquer* strategy to facilitate the construction of unified intermediate representation for the Java-side code and JavaScript-side code in React Native Android apps. REUNIFY focus on the Java-side code compiled into *Dalvik bytecode*. As mentioned in Section II, The JavaScript-side code in one React Native Android app can be either JavaScript code or Hermes bytecode depending on the JavaScript engine. As shown in Figure 3, for JavaScript code, REUNIFY uses *hermesc* (i.e. the Hermes compiler) that can compile JavaScript to Hermes bytecode but does not execute it. REUNIFY implements a front-end that can further transform readable assembly language disassembled by *hbctool* into the intermediate language of Soot.

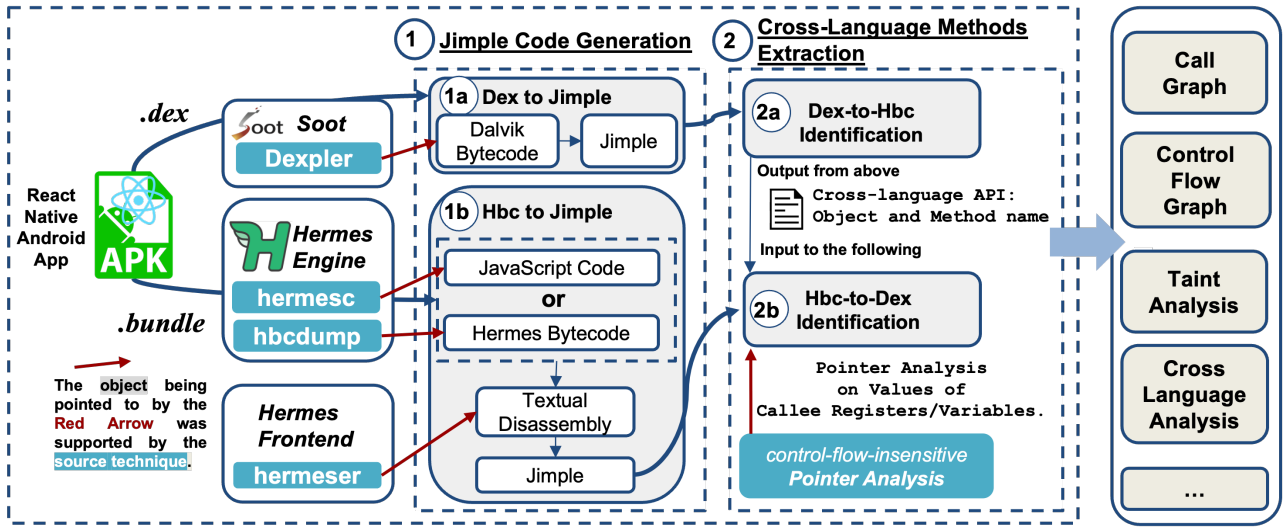

Fig. 3. Overview of ReuNify.

```
1   // JavaScript format
2   console.log("the String value longer than 16")
3
4   // Hermes Bytecode format
5   FILE HEADER: ...
6   FUNCTION HEADER TABLE: ...
7   STRING TABLE&STORAGE: ...
8
9   Function<global>(1 params, 11 registers, 0 symbols):
10  Offset in debug table: source 0x0000, lexical 0x0000
11      GetGlobalObject    r0
12      TryGetById         r2, r0, 1, "console"
13      GetByIdShort       r1, r2, 2, "log"
14      LoadConstString    r0, "the String value "...
15      Call2              r0, r1, r2, r0
16      Ret                r0
17
18  Debug Information: ...
```

Fig. 4. JavaScript code and the corresponding Hermes bytecode.

As shown in Figure 3, ①a the Dalvik bytecode files within the React Native Android app are transformed into Jimple by *Dexpler* [24] that is a front-end dealing with Dalvik bytecode, and has been integrated into Soot as one module. In the implementation in sub-step ①b of Figure 3, tool kits (including *hermesc* and *hbcdump*) of the Hermes engine converts the Hermes bytecode or JavaScript into a textual disassembly which can be further parsed and converted into Jimple code by the front-end, *hermeser*, proposed in REUNIFY.

In a typical analysis case, Soot is launched by specifying the target directory as a parameter. This directory contains the program (one *.apk* in this example) for analyzing. First, the *main()* method of the Main class is executed. It calls *Scene.loadNecessaryClasses()* where Soot locates the specified source code files (*.bundle* file for JS-side React-Native code in this example) from the input *.apk* file by *SourceLocator.v().getClassesUnder(path)*. Second, *HbcClassSource*, is implemented as a module inside Soot framework to create a *SootClass* from the corresponding disassembled Hermes bytecode. When the resolver has a reference to a *ClassSource* (*HbcClassSource*), it calls resolve() on it. *SootMethod*s are then created, and *MethodSource*s (corresponding to the information from the function in disassembled Hermes bytecode) are distributed for each *SootMethod*. When a Hermes bytecode method is stored into *MethodSource*, its opcode instructions are organized into blocks that can link to each other through the control flow. During the solving of each *SootMethod*, the Jimple instruction would be created for opcode instructions within all blocks, then the *jump* between each block can be connected. So that the generated Jimple code keep the same control flow with the Hermes bytecode.

The parts of the REUNIFY that translate one representation to another are inherently complex because they require the understanding of the semantics of both representations. In the following, we detail some challenges in transformation from Hermes bytecode to Jimple.

**Challenges related to Hermes bytecode disassembler:** Challenge ❶ Due to the evolution of the React Native framework, each version of React Native can only work with the compatible version of the Hermes engine. Until April 2023, there had been 42 versions of Hermes engine released for different version of React Native framework. Challenge ❷ The disassembled file from various versions of Hermes bytecode disassembler, *hbcdump*, can be different. Challenge ❸ The Hermes' build-in disassembler, *hbcdump*, does not reveal the full value of a String and Integer in the textual disassembly if their length exceeds 16 (as shown in line 14 in Figure 4, the String value is not fully disclosed in the disassembled Hermes

```
1  public class HermesByteCode {
2      public static JavaScript.Function.HermesByteode.global.JavaScript.FunctionOutput global(JavaScript.Object){
3          JavaScript.FunctionOutput r0;
4          Hbc.GlobalObject.console.log r1;
5          Hbc.GlobalObject.console r2;
6          JavaScript.Parameter_0 arg0;
7          arg0 := @parameter0: JavaScript.Parameter_0;
8          r0 = staticinvoke <Hbc.Opcode: Hbc.GlobalObject GetGlobalObject()>();
9          r2 = staticinvoke <Hbc.Opcode: Hbc.GlobalObject.console hbcGet(JavaScript.0bject,JavaScript,Number,JavaScript.String)>(r0, 1,"console");
10         r1 = staticinvoke <Hbc.Opcode: Hbc.GlobaObject.console.log hbcGet(JavaScript.0bject, JavaScript.Number, JavaScript.String)>(r2 , 2 , "log");
11         r0 = staticinvoke <Hbc.0pcode: JavaScripiString LoadConstString(JavaScript.String)>("the Stringvalue longer than 16");
12         r0 = staticinvoke <Hbc.GlobalObject.console: Hbc.GlobalObject.console.log.JavaScript.FunctionOutout log(JavaScript.0bject)>(r2, r0);
13         return r0;
14     }
15 }
```

Fig. 5. Jimple Code generated from the JavaScript code or Hermes Bytecode in Figure 4.

bytecode.).

**Challenges related to the program representation:** Unlike the source code or JVM bytecode, all instructions in Hermes bytecode are stored in functions that are as a sequential pile. Challenge ❹ In contrast to Java, Hermes bytecode does not use function names, parameters, and return values as signatures for function identification. In the Hermes bytecode representation, there may be functions with duplicate names, unnamed functions, and invocations of the same function with different numbers of arguments. Challenge ❺ In Hermes bytecode, the Function is considered the *First-class Object* similar to JavaScript, i.e., they can be manipulated like any other data type or variable, which enables functions to be passed as arguments, returned as values, and stored in registers. Moreover, all function invocation instructions occur on a register where a function was assigned before the invocation instruction was executed. As shown line 15 in Figure 4, *Call2 r0, r1, r2, r0* invoke the function store at register *r1* with arguments including *r2* and *r0*, and finally return value would be stored at and overwrite the value at *r0*. Also, as the dynamic feature of JavaScript, the type of each register in Hermes bytecode is not stable. As shown in Figure 4, the value of a register, *r0*, has been assigned by three instructions including *GetGlobalObject*, *LoadConstString*, and *Call2*, respectively.

**Solutions.** For challenges ❶, ❷, and ❸, we dedicated considerable engineering efforts to scrutinize each version of the source code of the 42 Hermes engines. For example, we engineered and build on the source code of 42 distinct versions of the Hermes engine to enable the complete display of String and Number values in the textual disassembly. Otherwise, only partial values are displayed by default when the length of Strings or Numbers exceeds 16. Subsequently, these built-in tools in Hermes are utilized in the Hermes-to-Jimple transformation, where we address the inconsistency in the textual disassemblies.

To address Challenge ❹, a unique SootMethod is created for each Hermes function with a specific signature before generating the corresponding Jimple statement. This is necessary to handle cases where there are duplicated function names, no function names, or uninclued parameter values used in the function body. The Hermes frontend tool, *hermser*, is used to extract information about each Hermes function from the textual disassembly and store it in an HbcMethod object. SootMethod is then created while taking into account the issues identified in the HbcMethod list. For example, if duplicated names (i.e., OneDuplicatedName) are detected in those *HbcMethod*s, a new name with one index (e.g., hermesDuplicatedFunction_OneDuplicatedName_5) would be used for the *SootMethod*.

To address Challenge ❺, a control-flow insensitive pointer analysis was implemented during the generation of Jimple statements. Hermes opcode instructions are mapped to Jimple statements, and registers are mapped to Jimple local variables. The type of local variables is generated and updated dynamically in the sequence of each Hermes instruction solved. As seen in Figure 5, from line 9 to line 11, the return value type changed from *Hbc.GlobalObject* to *Hbc.GlobalObject.console.log*. A method invocation then occurs on line 13 using *Hbc.GlobalObject.console* as the class name and *log* as the method name. This allows the method *console.log()* to be inferred in a fixed pattern. Each Hermes opcode instruction is mapped to a corresponding (or a group of) Jimple statements. Most Hermes opcode instructions would be generated as *staticinvoke* statement with *Hbc.Opcode* as the class name, Hermes opcode value as the method name. The return value in the method signature would be used for recording and tracking the dynamic type of return value. A comprehensive mapping can be found in REUNIFY's open-sourced project.

### B. Cross-Language Methods Extraction

In the previous step, we transform the JavaScript-side code of React Native apps into the intermediate representation in Soot, specifically *Jimple*, to facilitate a more thorough static analysis. However, due to React Native apps behaving native functionalities through the invocation of Java-side functions, the significance of Java-side code within these apps cannot be overlooked. In fact, within the REUNIFY framework, the analysis of the Java-side code is encompassed. The cross-language invocation on each side is not naturally connected to each other. To be able to extract the cross-language invocations are also an essential step toward whole program analysis.

**Step 2a: Dalvik-to-Hermes Invocation Extraction.** As mentioned in Section II, *React Native* is gradually replacing the legacy Architecture with the New Architecture. The underlying mechanism for cross-language communication have changed from *bridge* to *JavaScript Interface* (JSI) which enables direct calls from JavaScript to native code without the need for a *bridge*. The implementation for developers to create Native API and Native Components is also changed with the update of Architecture. *Native Module* and *Native Components* are the established technologies utilized in the legacy architecture. They will be deprecated in the future once the New Architecture becomes stable. The New Architecture uses *Turbo Native Module* and *Fabric Native Components* to achieve similar results [10]. In this case, we take Turbo Native Module as one example to explain step ②a of REUNIFY. (The intuition is that the identification process for *Native Module* is similar and easier than the *Turbo Native Module*.)

This step is performed over 5 sub-steps: ❶ Analyzing class hierarchy to record classes that extend *ReactContextBaseJavaModule* and also implement both *ReactModuleWithSpec* and *TurboModule* as shown in the line 2 and line 3 of Figure 6. ❷ Records the name for the method with *@ReactMethod* annotation. (as shown in the line 10 of Figure 6) ❸ Track class hierarchy to detect the classes that extend the classes recorded in sub-step 1 (as shown in the line 14 of Figure 6). ❹ Go through the methods in the class recorded in sub-step 3, and retrieve out the methods that overwrite the methods recorded in sub-step 2 (as shown in the line 27 of Figure 6). ❺ Retrieve the method with the sub-signature, *java.lang.String getName()* (as shown in line 22 of Figure 5), and further extract the return value of this method (e.g., the return value is *Calendar* at line 15 of Figure 6) as the Module API name.

The aforementioned procedure showcases REUNIFY's approach for the Dalvik-to-Hermes Identification within the New Architecture (i.e., *Turbo Native Module*). Implementing the Dalvik-to-Hermes Identification within the Old Architecture (i.e., Native Module) is less challenging than the aforementioned process. A process resembling sub-step ❶ is essential to discover the class that encapsulates the *Native Module*, where methods annotated with *@ReactMethod* would be considered as Module API methods. Subsequently, sub-step ❺ can be carried out to determine the Module API name in the same class. A similar approach can be used to identify cross-language communication on the Java side for *Native Components* and *Fabric Native Components*, by tracking the method annotated with either *@ReactProp* or *@ReactPropGroup*. However, due to space limitations, we cannot provide all the technical details here. For a more comprehensive understanding, please refer to REUNIFY's open-source project.

**Step 2b: Hermes-to-Dalvik Invocation Extraction.** The

```
1  public abstract class CalendarModuleSpec
2      extends ReactContextBaseJavaModule
3      implements ReactModuleWithSpec, TurboModule {
4      public CalendarModuleSpec(ReactApplicationContext rContext)
5      {
6          super(rContext);
7      }
8
9      @DoNotStrip
10     @ReactMethod
11     public abstract void createCalendarEvent(int i1, int i2, String str);
12 }
13
14 public class CalendarModule extends CalendarModuleSpec {
15     public static final String RN_CLASS ="Calendar";
16
17     public CalendarModule(ReactApplicationContext rContext) {
18         super(rContext)
19     };
20
21     @Override
22     public String getName() {
23         return RN_CLASS;
24     }
25
26     @Override
27     public void createCalendarEvent(int i1, int i2, String str) {
28         Intent intent = new Intent("android.intent.action.INSERT");
29         intent.setData(CalendarContract.Events.CONTENT_URI);
30         intent.putExtra("title" , str);
31         getReactApplicationContext().startActivity(intent);
32     }
33 }
```

Fig. 6. Module API registration example in New Architecture of React Native.

Module API name and methods name retrieved from the Step ②a would be used as the identifier for the cross-language invocation on the JavaScript side. Compared to Java code analysis, pointer analysis is more challenging in Hermes bytecode due to the language's dynamic feature, as Hermes bytecode is compiled from JavaScript. This means that register values are not determined until runtime, which potentially leads to instability of the value in function invocation's callee registers and complicates analysis. To address this, we use a control-flow-insensitive technique to track the value stored in a register (variable). As seen in Figure 5, from line 9 to line 11, the return value type changed from *Hbc.GlobalObject* to *Hbc.GlobalObject.console.log*.

In the process of Hermes-to-Jimple transformation, all the registers that are used as callee of function invocations are recorded. The Module API names and method names retrieved from step ②a are used as a filter to detect the Hermes-to-Dalvik invocation. To implement the cross-language invocation for the Java-side code from the JavaScript side, one object name will be used on the JavaScript side to access the object that is exposed from the Java-side code. In the example in Figure 6, the value, *Calendar*, which is retrieved by the sub-step ❺ at Step ②a, is the object name exposed to the JavaScript side code. To access the method wrapped into the exposed cross-language object, the method name would be used to retrieve the value (Java-side function) stored into key-value pair. The method name, *createCalendarEvent*, will be used to refer the function at line 27 at Figure 6. To implement invocation at *hbc*, the Hermes opcode instruction for function invocation is used with the callee register. By comparing the value of each callee register with the Module API and function name, the potential Hermes-to-Dalvik Invocation can be retrieved.

The accuracy and precision of Hermes-to-Dalvik Invocation Analysis rely on effective pointer analysis of function invocation's callee registers. However, the use of *First-class Objects* for all Hermes functions adds complexity to comprehending program behavior in static analysis. This complication is especially notable in intricate systems like React Native and during code transformations using JavaScript bundlers.

## IV. EVALUATION

In this section, we commence by undertaking a preliminary study to explore the extent of React Native's utilization. Subsequently, we delve into the following research questions to gauge the significance of our contributions:

- **RQ1:** How well does REUNIFY enhance Soot-based static analysis on React Native Android Apps?
- **RQ2:** Can REUNIFY reveal previously unreachable sensitive data leaks in React Native Android Apps?

We ran all of our experiments on a Linux server with Intel (R) Core (TM) i9-9920X CPU @ 3.50GHz and 64 GB RAM.

### A. Preliminary Study

We first conducted a preliminary study to explore the utilization of the React Native framework across a spectrum of Android apps, encompassing both popular and potentially malicious applications.

**Dataset:** To create a dataset of popular Android apps, we began by gathering a list of 15,854 Android apps from ANDROIDRANK [25]. This list included the top 500 apps for each of the 32 app categories available on Google Play. We then downloaded the latest version of 14,874 out of 15,854 of these apps from AndroZoo [26]. The remaining 980 apps were not available for download.

In addition, we obtained a dataset of 60,618 malware apps from VirusShare [27], which included Android malware apps collected by VirusShare in 2022. We also gathered 67,135 malicious apps from AndroZoo. We consider an app to be malicious if at least 10 antivirus engines in VirusTotal had flagged it.

**Study Design:** The React Native framework is developed using multiple programming languages, including Java, C++, JavaScript, Objective-C, and others [28]. The framework code is typically included in the release build to ensure proper app functionality. To gauge the extent of React Native framework adoption in Android apps, we conducted a preliminary study in which we examined the APK file of each app for the presence of the Java package, *com.facebook.react*. It is noteworthy that code Obfuscation will not affect this package name [29]. However, since React Native Java libraries can be included in apps as a single library but not actually used, we also verified the presence of a bundle file for the JavaScript-side code within each app's APK file. Analyzing the type of a bundle file can reveal whether it contains Hermes bytecode or JavaScript code.

TABLE I
JAVASCRIPT-CODE FORMAT IN MOST POPULAR APPS AND MALWARE APPS

| Category | Hermes Bytecode | JavaScript | Total |
|---|---|---|---|
| Popular | 494 | 574 | 1 068 |
| Malware | 28 | 413 | 441 |
| **Total** | 522 | 987 | 1 509 |

**Results:** Our empirical study indicates that **1,068 apps, accounting for 7.2% of those 14,874 most popular apps collected from AndroZoo**, were developed using the React Native framework. Of these React Native Android apps, 494 (46.3%) utilized the Hermes engine as the JavaScript runtime and compiled the JavaScript into Hermes bytecode. In contrast, among the 60,618 malware collected from VirusShare, **there were 441 apps developed with the React Native framework**. Out of these 441 React Native Android malware apps, only 28 of them used the Hermes engine.

> Within the selection of the 14,874 most popular Android applications, approximately 7.2% have been created using the React Native framework. The presence of malware has extended to encompass React Native applications as well. Furthermore, the employment of the Hermes engine exhibits lower frequency among malware apps in comparison to its prevalence within popular applications.

### B. RQ1: How well does REUNIFY enhance Soot-based static analysis on React Native Android Apps?

Our objective with this RQ is to understand how REUNIFY enables the static analysis on React Native Android Apps. We evaluate REUNIFY on those 494 Hermes engine-enabled apps out of the 1,068 most popular React Native Android Apps from two perspectives: ❶ the number of generated Jimple Code, ❷ the number of identified Dalvik-to-Hermes invocation. Since the implementation of Hermes engine impacts the volume of code implemented in React Native framework in Android apps [30], we focused our analysis on popular apps that adopted the Hermes engine for a fair comparison. Furthermore, since the Hermes engine has become the default engine of React Native, our findings offer more valuable insights into the current state of React Native Android apps using the Hermes engine. To further assess the practicality of REUNIFY, we utilized FlowDroid to generate callgraphs for 1,068 popular React Native apps and 441 React Native malware apps, and compared the ❸ size of the callgraphs before and after integrating REUNIFY. A diverse set (including both popular apps and malware apps) of React Native apps can further prove the effectiveness of REUNIFY.

TABLE II
AVERAGE NUMBER OF NATIVE MODULE API AND NATIVE COMPONENT UI

| Category | Apps | Native | | | |
|---|---|---|---|---|---|
| | | Module API | Module API Methods | Component UI | Component UI Methods |
| Popular | 494 | 92 | 532 | 55 | 489 |
| RN Toy App | 1 | 51 | 213 | 22 | 365 |

TABLE III
AVERAGE NUMBER OF VOLUME OF CODE

| Category | # apps | Soot without ReuNify | | Soot with ReuNify | | Difference | |
|---|---|---|---|---|---|---|---|
| | | # Methods | # LOC | # Methods | # LOC | # Added Methods | # Added LOC |
| Popular | 494 | 132 093 | 1 697 294 | 162 195 | 2 879 754 | 30 102 (+22.79%) | 1 182 460 (+69.67%) |
| RN Toy App | 1 | 43 005 | 476 859 | 47 209 | 632 880 | 4 204 (+9.78%) | 156 021 (+32.72%) |

TABLE IV
AVERAGE NUMBERS OF NODES AND EDGES BEFORE AND AFTER REUNIFY ON 1,007 MOST POPULAR APPS AND 421 MALWARE APPS

| Category | # apps | without ReuNify | | with ReuNify | | Difference | |
|---|---|---|---|---|---|---|---|
| | | # Nodes | # Edges | # Nodes | # Edges | # Added Nodes | # Added Edges |
| Popular Apps | 1 007 | 9 206 | 70 344 | 16 940 | 102 830 | 7 734 (+84.01%) | 32 486 (+46.18%) |
| Malware Apps | 421 | 6 465 | 36 572 | 9 824 | 48 460 | 3 359 (+51.96%) | 11 888 (+32.51%) |

**Volumn of Jimple Code:** The quantity and quality of static analysis results produced by Soot's framework are heavily reliant on the availability of Jimple code. With REUNIFY's *hermeser* integrated into Soot framework, an additional *class* for Hermes bytecode is created. This class comprises an average of 30,102 SootMethods with 1,182,460 lines of Jimple code. According to Table III, **with the augmentation of REUNIFY's *hermeser*, there are 70% more Jimple statements generated compared with 1,697,294 lines of code generated by Soot.**

REUNIFY successfully generated the additional Jimple statement for 452 out of 494 apps. The unsuccessful cases were due to the customizable nature of the bundle name [31], which made it difficult to locate the JavaScript-side bundle file. To improve the reliability of the analysis, future work should focus on developing more robust techniques for locating bundle files. For the apps with located bundle files, all of them were successfully transformed into Jimple code from Hermes bytecode. Moreover, all *SootMethod*s generated by REUNIFY's *hermeser* passed Soot's body validation (<*soot.jimple.JimpleBody: void validate()*> [32] in Soot), indicating that the generated Jimple code is valid in the Soot framework. This allows for additional Soot-based analysis on Hermes bytecode.

**Number of Hermes-to-Dalvik invocation:** React Native enables accessing methods on the Java side from the JavaScript side. As shown in Table II, React Native apps have an average of 93 Native Module APIs, which contain 569 methods accessible to JavaScript code, and 52 Native React Components comprising 477 methods for setting UI attributes. As shown in Figure 2, the Native Module APIs and Components can be sourced from the React Native framework, third-party libraries, or the developer's own implementation. To determine the extent of Hermes-to-Dalvik invocations coming from sources beyond the Core Module APIs and Core Components, we build a Toy app from the project (React Native CLI Quickstart [33]) in React Native version 0.71. This Toy app only includes the Core Module APIs and Core Components without any developer's code or third-party library. According to Table II, **the most popular React Native Android apps have over twice the number of Native Module API methods (532 methods) compared to the React Native Toy app (213 methods)** using React Native version 0.71. With the use of Native Module API and Native Components, more powerful functionalities (in terms of performance and access to system resources) can be exposed to the JavaScript side. It is customary to involve extra Native Module APIs and Native Components while developing a React Native Android app.

**Size of Callgraph:** In static analysis models, callgraph is a crucial component as it offers a complete perspective of the program's behaviour. To evaluate the effectiveness of REUNIFY in generating callgraphs, we compared the size of callgraphs produced by FlowDroid with and without the augmentation of REUNIFY, for both popular and malicious React Native Android Apps. Out of the 1,068 most popular React Native apps and 441 React Native malware apps, callgraphs get generated successfully on 1,007 and 421 apps respectively, with or without the use of REUNIFY. Nonetheless, in some cases, due to time limitations or obfuscation techniques, Callgraph failed to be generated on 61 popular apps and 20 malware apps.

We first report the average number of nodes (i.e., the number of methods) and edges (i.e., the number of potential invocations) in the callgraphs obtained before and after having applied REUNIFY. The call-graph augmentations introduced by REUNIFY can be seen in Table IV, where the number of apps affected by the changes is represented by the # apps column. We observe that all apps' callgraphs are enlarged by the use of REUNIFY (1,007 and 421 for popular and malware apps, respectively). Additionally, we notice that the number of nodes and edges uncovered with REUNIFY is higher for popular apps than for malware apps: 7,734 vs 3,359 on average per app for nodes and 32,486 vs 11,888 for edges. This highlights that **traditional static analyzers that do not consider the executable code in React Native apps miss a substantial number of nodes and edges** in their call graphs.

By considering the mechanism of React Native, REUNIFY can identify previously unreachable Java methods that are now reachable. The number of such previously unreachable methods is highly correlated with the number of Hermes-to-Dalvik invocations. The discovery of newly reachable nodes is significant because it allows static analyzers to avoid treating them as dead code.

> **Answer to RQ1:** Soot tends to miss a significant portion of executable code when analyzing React Native Android apps. However, by converting Hermes bytecode to Jimple, there is a 70% increase in the number of lines of Jimple code in Soot. Taking into account the React Native mechanism on the Java side, popular apps experience an increase of approximately 84% in new nodes for callgraph, while malware apps experience an increase of around 52% in nodes for callgraph.

*C. RQ2: How effective is REUNIFY in finding sensitive data leaks in React Native Android Apps?*

In this research question, we demonstrate the capability of REUNIFY in finding potential privacy leaks in real-world React Native Android apps.

**Experimental setup:** In order to evaluate the effectiveness of REUNIFY in finding privacy leaks, we conducted experiments on both popular apps and malware to demonstrate its effectiveness. Specifically, we tested REUNIFY on 1,068 of the most popular React Native Android apps, as well as 441 React Native malware instances detected in the year 2022 and sourced from VirusShare[27]. In order to ensure a fair comparison, we utilized the default sources and sinks provided by FlowDroid. However, it should be noted that REUNIFY supports custom sources and sinks tailored to specific needs and interests, such as those pertaining to JavaScript. Sources and sinks in the context of privacy leaks refer to the entry and exit points in an app's code where data can enter and leave the system. FlowDroid is capable of identifying data flows from sensitive sources to potentially unsafe sinks. It is important to keep in mind that dataflow analysis can be both time and memory intensive, and therefore, for each app, we set a maximum time limit of 30 minutes for FlowDroid to complete its analysis.

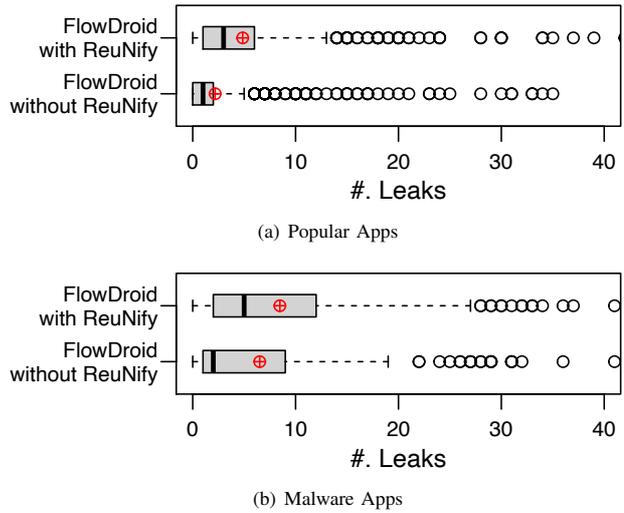

(a) Popular Apps

(b) Malware Apps

Fig. 7. Distribution of the number of leaks detected by FlowDroid with and without REUNIFY

**Findings:** FlowDroid was executed successfully on 1,007 out of the 1,068 most popular apps and on 421 out of the 441 malware apps, with or without REUNIFY augmentation. However, due to time constraints or obfuscation techniques, FlowDroid failed to run on 61 of the most popular apps and 20 of the malware apps. **In total, applying REUNIFY resulted in the detection of 2,690 (4,892 − 2,202) additional privacy leaks for popular apps and 827 (3,576 − 2,749) additional privacy leaks for malware apps, respectively.** The average number of leaks is indicated by the ⊕ labels in Figure 7(a) and Figure 7(b), respectively. On average, Figure 7(a) indicates that by incorporating REUNIFY, an extra 2 privacy leaks (totaling 4 leaks) were identified in popular apps compared to only running FlowDroid, which could detect only 2 leaks. Similarly, as shown in Figure 7(b), with the augmentation of REUNIFY, an additional 2 privacy leaks were detected on average, making a total of 8 leaks, compared to only running FlowDroid (i.e., 6 leaks) for those malware apps. It is not surprising that more leaks are detected from malware apps than popular benign apps, as the number of leaks is highly reflective of potential issues in an app.

**Types of newly detected privacy leaks:** After identifying privacy leaks additionally discovered by REUNIFY, we further categorize the sources and sinks according to SuSi's classification [34] to facilitate understanding of each privacy leak. For any sources or sinks that were not classified, we manually assigned categories based on the functionality of their classes and methods. Among them, the most common sink type was the *Replace* sink, represented by the method <*java.lang.String: java.lang.String*

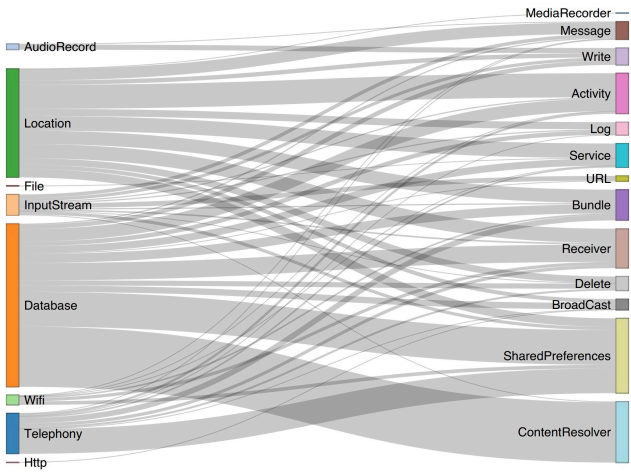

Fig. 8. Sankey diagram of all newly detected privacy leaks.

*replace(java.lang.CharSequence,java.lang.CharSequence)>*. The method, *replace*, is frequently used to substitute a particular sequence of characters in a string with another sequence of characters. However, if sensitive data (e.g., user credentials) is included in either the original or replacement character sequences, this information can be inadvertently leaked. We found that for both popular and malware apps, the most common type of leaked information was data stored in the database. The second most common type of leaked information for popular apps was Wi-Fi-related information including Service Set Identifier (SSID) and MacAddress. For malware apps, the second most common leaked type of source information was telephony information, including Device Id, Line1Number (phone number of the device's SIM card), subscriber ID, and SimSerialNumber. For both popular and malware apps, more than 98% of sources that leaked from the method, *replace*, come from the top two most common sources as described above.

To better comprehend and visualize additional privacy leaks discovered by REUNIFY, we have created a Sankey diagram (Figure 8) that includes newly detected leaks for both popular and malware apps while excluding the leaks with *replace* as sinks. It can be observed from Figure 8 that the primary sources of privacy leaks are Database, Location, and Telephony. The sensitive information is predominantly leaked to SharedPreferences, ContentResolver, and Activity. In fact, our analysis shows that the use of REUNIFY resulted in a significant increase in the number of detected sensitive data leaks for both popular Android apps and malware.

> **Answer to RQ2:** REUNIFY is effective for identifying data leaks that were previously unseen. Specifically, on average, 2 additional potential leaks can be detected in both popular apps and malware.

## V. THREATS TO VALIDITY

**Analyzing C++ code.** JavaScriptInterface (JSI) is a module in React Native's new architecture that can enable C++ code to be invoked from the JavaScript side, which could affect the comprehensiveness of our study. However, the usage of JSI and C++ is still experimental, and the implementation of C++ is being gradually automated by the Codegen module [35] in the React Native framework. Since C++ code is not as prevalent as Java code at present, REUNIFY plans to gradually incorporate support for C++ code in the future

**Transforming Hermes bytecode to Jimple.** The Hermes engine has been used as the default engine for React Native applications. The JavaScript code used in the development of React Native applications is compiled into Hermes bytecode. One feature of the REUNIFY framework is its capability to abstract different versions of Hermes bytecode into the Jimple statements. Due to the dynamic and intricate program representation of Hermes bytecode, the Jimple code generated from it has some limitations when it comes to handling complicated point analysis, such as the taint analysis. In this study, REUNIFY has interpreted the Hermes opcodes that pertain to conditional branching and function declaration during the transforming process. Certain static techniques, such as Single Static Assignment (SSA) and intraprocedural Control Flow Graph (CFG), can be implemented on the produced Jimple code. In our prospective endeavors, we would like to better interpret Hermes opcodes into Jimple, hence facilitating the execution of more complex static analysis.

**Method Invocation from the JavaScript Side.** One major limitation of REUNIFY is its inability to accurately model the behavior of functions in Hermes bytecode, which is due to the lack of a point analysis technique for Hermes bytecode. The points-to analysis in Soot cannot directly work on the Jimple code generated from Hermes bytecode, mainly due to the dynamic-language feature and the presentation of Hermes bytecode. However, we present a solution that utilizes a control-flow-insensitive technique to infer the type of the register value, which can identify some invocations for the Java-side code and recover build-in API methods (e.g., *console.log()*, *alert()*, *JSON.parse()*, etc.). Nonetheless, it remains a challenge to verify if the Hermes-to-Dalvik identification has yielded correct links. One possible way to verify this would be to execute the code section to trigger the native code and ensure that the correct information is yielded by Hermes-to-Dalvik indentification. However, this is beyond the scope of this study. Therefore, we have made the hypothesis that the correct results are yielded from Hermes-to-Dalvik identification.

## VI. RELATED WORK

**Java-based Android Apps Analysis.** Li et al [6] provide a comprehensive survey of Android apps, focusing on static analysis approaches. Different static analysis approaches are utilized to detect compatibility issues [35, 36, 37, 38, 39] and other functional or non-functional faults [40, 41, 42, 43, 44, 45]. Moreover, static analysis can be leveraged to collect information in apps towards improving dynamic testing approaches [46, 47, 48, 49]. The popular artifacts adopted by current researchers are MalloDroid by Fahl et al. [50], which detects improper use of transport layer security in apps; FlowDroid by Arzt et al. [19], which is able to find privacy leaks by inspecting illicit information flow; and IccTA by Li et al. [21], which extends FlowDroid by accounting for inter-component privacy leaks. Instead of focusing on Java-based Android apps analysis, our work has taken a step forward by proposing an approach to take an additional programming language, Hermes bytecode/JavaScript (used in React Native Android Apps), into consideration. We expect to provide the community with a readily usable framework, which enables researchers and practitioners to complete their analyses on React Native Android Apps.

**Analysis of Multiple Languages in Android App.** The research emphasis has been on analyzing languages used in Android Apps beyond just Java, and also on conducting cross-language analysis. Lee et al. [18] analysed the inter-communication between Android Java and JavaScript and presented the framework, HybriDroid, to detect bugs and information leaks in hybrid apps. However, HybriDroid is Android version sensitive and only focuses on the bridge communication between Android Java and JavaScript (the other communication approach is callback communication). Alam et al. [51], in 2016, proposed DroidNative, which can perform Android malware detection considering both the bytecode and the native code. What's more, NDroid [52], TaintArt [53], and PolyCruise [54] were proposed for dynamic taint analysis so as to track sensitive information flows. JN-SAF and Jucify [20, 55] are also proposed as an inter-language static analysis framework to detect sensitive data leaks in Android apps. The Jimple statements produced by *Jucify* are insufficient and unable to capture the complete implementations of the native functions, which poses a challenge in commencing further research (inter-procedural analysis) on the native code for whole program analysis. All the aforementioned tools, however, are task-specific. They also, typically, perform their analyses separately for bytecode and native code, and later merge the outputs to present unified analysis results. In contrast, REUNIFY is proposed to unify the representation before task-oriented analyses, which empowers popular analysis pipelines to be directly adopted on the output of REUNIFY.

## VII. CONCLUSION

As the use of React Native Framework continues to increase in the development of Android applications, it is no longer feasible to disregard it for static analysis. Our approach, RE-UNIFY, considers the mechanism of React Native framework. By manipulating code at the level of Soot IR, REUNIFY effectively augment the current Soot-based static analysis tools for React Native App analysis. Our evaluation of the most popular real-world React Native Android programs shows that REUNIFY significantly enhances static analysis and call graph comprehensiveness. Furthermore, when we ran FlowDroid with REUNIFY, we discovered an average of two additional privacy leakages for 1,007 most popular React Native Android Apps. We are confident that these findings establish our approach as a necessary improvement to the well-established static analysis for React Native Android programs.

ACKNOWLEDGEMENTS

Grundy is supported by ARC Laureate Fellowship FL190100035.